\documentclass[twoside,fleqn]{article}
\usepackage{espcrc2,latexsym}

\usepackage{epsfig}

\newcommand{\Nf}{N_{\!f}}

\newcommand{\ovr}{\over}

\newcommand{\<}{\langle}
\renewcommand{\>}{\rangle}

\newcommand{\et}{\eta}

\newcommand{\ka}{\kappa}

\newcommand{\rh}{\rho}

\newcommand{\beq}{\begin{equation}}
\newcommand{\eeq}{\end{equation}}
\newcommand{\bdm}{\begin{displaymath}}
\newcommand{\edm}{\end{displaymath}}
\newcommand{\bea}{\begin{eqnarray}}
\newcommand{\eea}{\end{eqnarray}}

\title{Topology and pion correlators -- a study in the $N_{\!f}\!=\!2$
Schwinger model%
}

\author{S. D\"urr
\address{Paul Scherrer Institut, Theory Group, 5232 Villigen PSI, Switzerland}%
}

\begin{document}


\begin{abstract}
I readdress the issue whether the topological charge of the gauge background
has an influence on a hadronic observable. To this end pion correlators in the
Schwinger model with 2 dynamical flavours are determined on subensembles with
a fixed topological charge. It turns out that the answer depends on a specific
function of the sea-quark mass and the box volume which is in close analogy
to the Leutwyler-Smilga parameter in full QCD.
\vspace*{-1mm}
\end{abstract}

\maketitle



\section{INTRODUCTION}

Does topology matter ? Or, more explicitly: Does the global topological charge
of a QCD configuration have an influence on a typical hadron correlator
determined on that background and does thus the charge distribution of an
ensemble affect physical measurements ?

In a first round, one has to distinguish between observables which relate to
the $U(1)_A$ issue and those which do not.
The mass of the $\et'$ is known to be sensitive, since it depends directly on
the distribution of topological charges via the explicit breaking of the
$U(1)_A$ symmetry through instantons.
On the other hand, standard observables like $M_\pi, M_K, M_\et, M_\rh$ or the
heavy-quark potential $V_{q\bar q}$ are generally known to be independent of
the index of the background.
Below, I address the latter category and the precise conditions (regarding the
four-volume $V$ and the sea-quark mass $m$) that have to be met for their
supposed insensitivity on topology to become true.

Leutwyler and Smilga have shown \cite{Leutwyler:1992yt} that for the specific
case of pionic observables much can be said on analytical grounds, if the
parameter\footnote{
$\Sigma\!=\!-\!\lim_{m\to 0}\lim_{V\to\infty}\langle\bar\psi\psi\rangle$
is the condensate in the chiral limit, $m$ is the actual (degenerate)
{\em sea\/}-quark mass; note that $\Sigma$ and $m$ are both scheme- and
scale-dependent, only the combination is an RG-invariant quantity.}
\beq
x\equiv V\Sigma m
\label{LSPdef}
\eeq
is known. Its role is to discriminate `small' from `large' boxes in the sense
that the magnitude of (\ref{LSPdef}) decides whether the system prefers to show
symmetry restoration or spontaneous symmetry breaking phenomena
($N_{\!f}\!\geq\!2$):
For $x\!\ll\!1$ the chiral symmetry is effectively restored and a good
description is in terms of quarks and gluons.
For $x\!\gg\!1$ the $SU(N_{\!f})_A$ symmetry is effectively broken (though the
box-volume is formally finite), meaning that pions represent appropriate
degrees of freedom.
Below it is important to keep in mind that the classification w.r.t.\
$x$ is {\em independent\/} of the standard classification w.r.t.\ the ratio
`pion correlation length to box size' ($1/M_\pi L$).

For $x\!\ll\!1$ the partition function is dominated by the topologically
trivial sector \cite{Leutwyler:1992yt}
\beq
Z_\nu \propto m^{N_{\!f}|\nu|}
\;,
\label{LSpeak}
\eeq
and one expects a clear {\em sectoral dependence\/} of all observables,
including the standard $M_\pi$.

In the opposite regime $x\!\gg\!1$ and with the auxiliary condition that the
quark masses are so light that the {\em sea\/}-pion substantially overlaps the
box ($1/\Lambda_\mathrm{eff}\!\ll\!L\!\ll\!1/M_\pi$), the path-integral in the
effective description is dominated by the constant mode~\cite{Gasser:1987ah},
the partition function gets broad \cite{Leutwyler:1992yt}
\beq
Z_\nu \propto e^{-{\nu^2\ovr2\<\nu^2\>}}
\;\;{\rm with}\;\;
\<\nu^2\>\!=\!{V\Sigma m\ovr N_{\!f}}
\;,
\label{LSgaussian}
\eeq
and as a result standard observables are expected to be {\em approximately
independent\/} of $\nu$.

From a lattice perspective, the main problem with the Leutwyler-Smilga (LS)
analysis is that it involves the wrong `auxiliary' condition in the large $x$
regime (one would like the pion to fit into the box rather than to spread
itself uniformly in three-space).
Furthermore, it would be desirable to investigate the issue directly at the
level of observables rather than to deduce their sensitivity on $\nu$ from that
of the partition function.


\begin{figure*}[t]
\epsfig{file=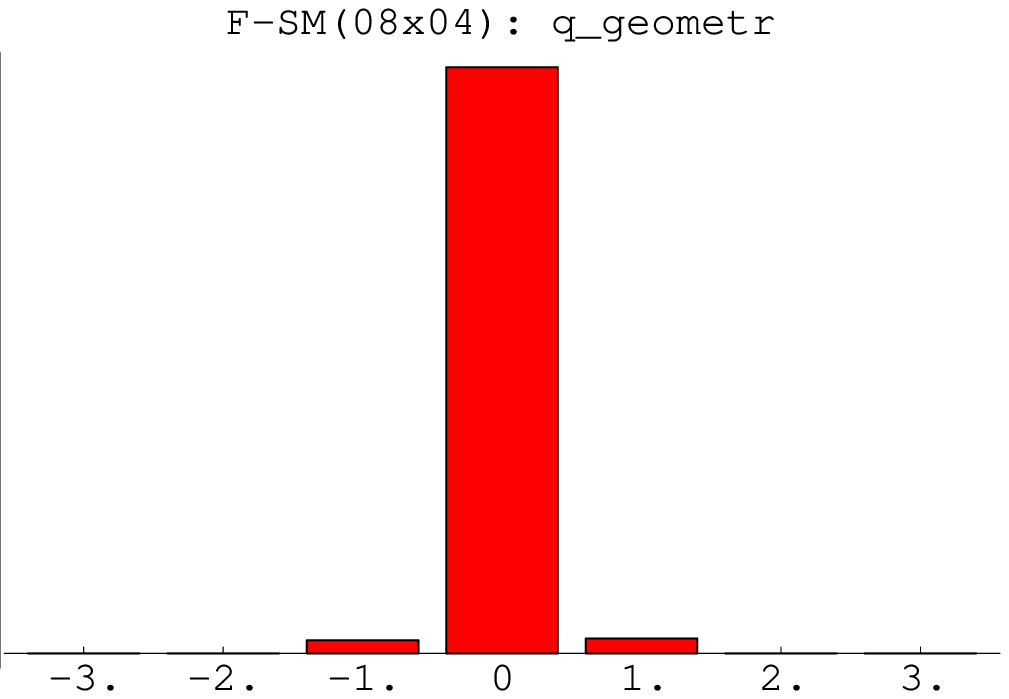,width=5.2cm}
\hspace*{0.1mm}
\epsfig{file=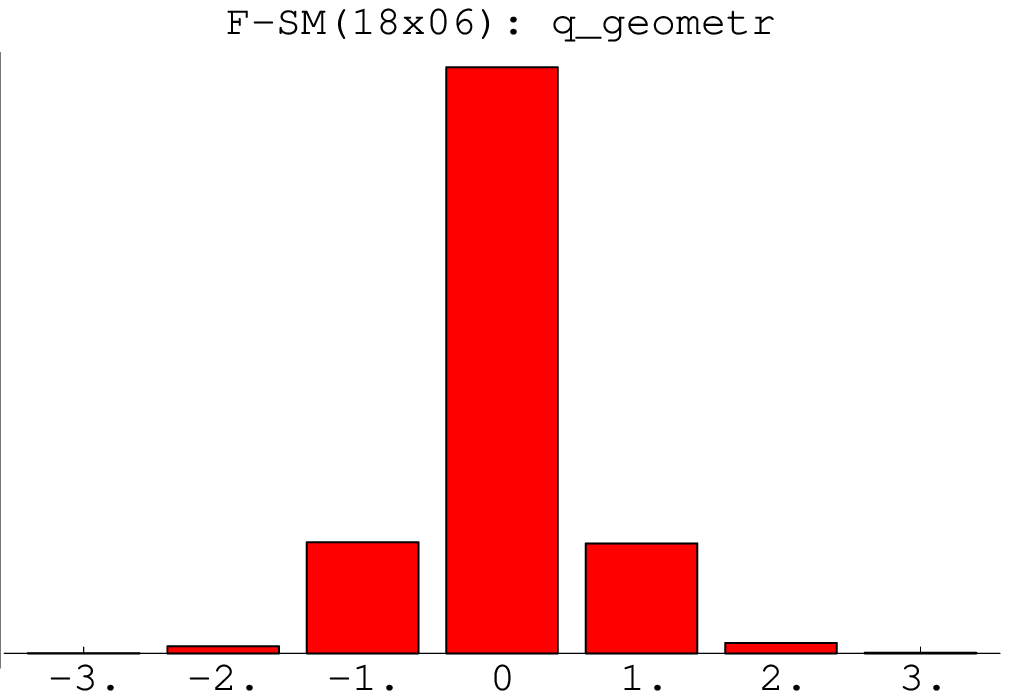,width=5.2cm}
\hspace*{0.1mm}
\epsfig{file=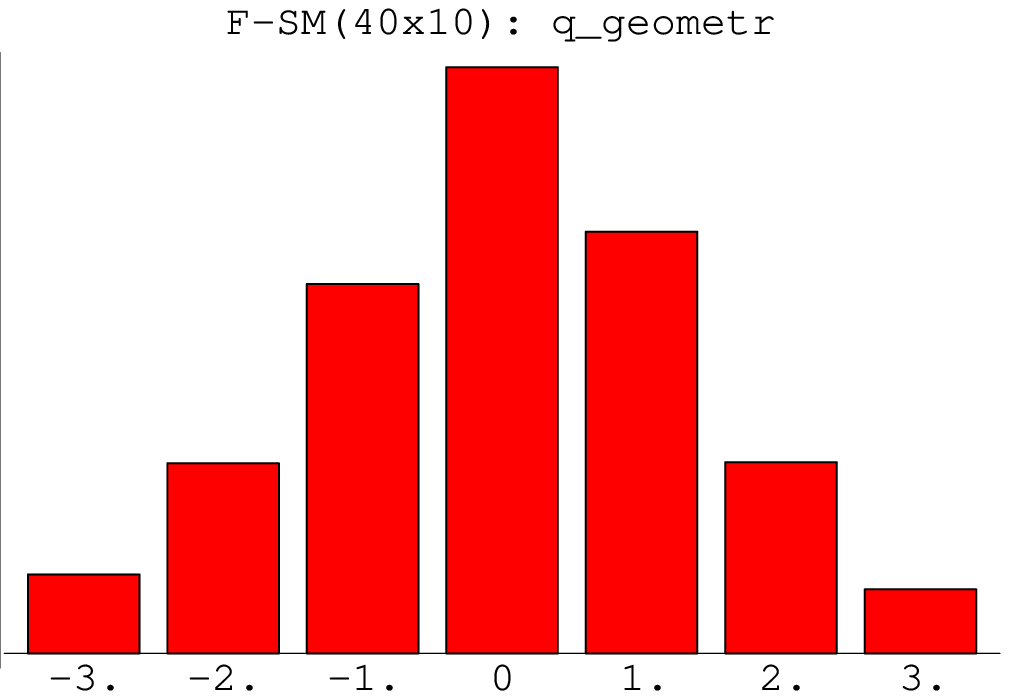,width=5.2cm}
\vspace*{-13mm}
\caption{Distribution of $\nu_{\rm geo}$ 
in the small ($8\!\times\!4$ lattice), intermediate ($18\!\times\!6$) and large
($40\!\times\!10$) Leutwyler-Smilga regimes, respectively [QED$_2$ data for
$\beta\!=\!3.4, m\!=\!0.09, N_{\!f}\!=\!2$ (staggered)].}
\vspace*{-2mm}
\end{figure*}

\section{SIMULATION DETAILS}

As the issue is peculiar to the dynamical theory, a pilot study in a suitable
model seems justified.
The massive multiflavour Schwinger model (QED$_2$ with $\Nf\!\geq\!2$) shares
the qualitative features needed \cite{Durr}.
The ensemble is generated with the Wilson gauge action
$S_{\rm gauge}\!=\!\beta\sum(1\!-\cos\theta_\Box)$ and a pair of staggered
fermions.
The plan is to compare the regimes $x\!\ll\!1, x\!\simeq\!1$, $x\!\gg\!1$
to each other using three dedicated simulations:
Working at fixed $\beta\!=\!1/g^2\!=\!3.4$ and $m\!=\!0.09$, the regimes are
represented by the volumes $V\!\!=\!8\!\times\!4,18\!\times\!6,40\!\times\!10$.
Given the approximate solution of the bosonized model, the LS parameter is
expected to take the values $x\!\simeq\!0.33,1.12,4.16$, respectively, and the
`pion' (pseudo-scalar iso-triplet) to have a mass $M_\pi\!\simeq\!0.329$ and
hence a correlation length $\xi_\pi\!\simeq\!3.04$ as to fit into the box (see
\cite{Durr} for details and \cite{SMlat} for references to and an assessment of
the bosonized solution).

A configuration is assigned an index only if the geometric
($\nu_{\rm geo}\!=\!{1\over2\pi}\sum\log U_\Box$) and the
field-theoretic definition ($\nu_{\rm fth}\!=\!\kappa\,\nu_{\rm nai}$,
$\nu_{\rm nai}\!=\!\sum\sin\theta_\Box$ with
$\kappa\!\simeq\!1/(1\!-\!\langle S_{\rm gauge}\rangle/\beta V$), after
rounding to the nearest integer, agree. 
Since this turned out to be the case for 99.9\%, 98.8\%, 88.1\% of the
configurations in the small/intermediate/large lattice, it means that for the
majority of configurations an assignment can be done {\em without cooling\/}.
Fig.\ 1 shows that the associate partition function obeys the LS prediction:
For $x\!\ll\!1$ it essentially consists of the charge zero contribution,
for $x\!\gg\!1$ it gets broad and gaussian -- even though the `auxiliary'
condition $M_\pi L\!\ll\!1$ in the LS analysis has been reversed into what is
usual on the lattice.


\section{SECTORAL PION PROPAGATORS}

\begin{figure*}[t]
\epsfig{file=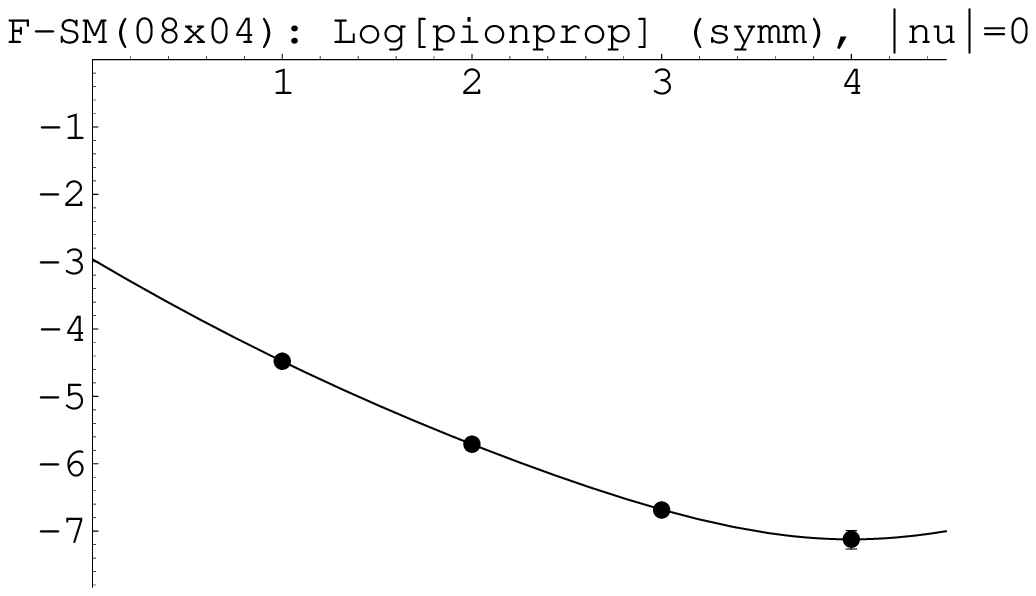,width=5.2cm,height=3.0cm}
\hspace*{0.1mm}
\epsfig{file=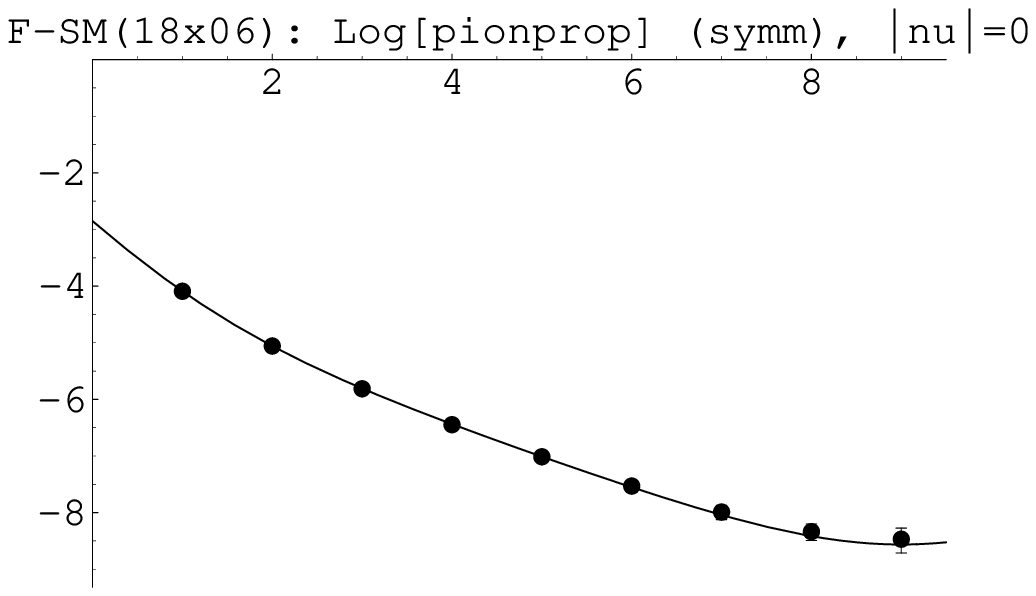,width=5.2cm,height=3.0cm}
\hspace*{0.1mm}
\epsfig{file=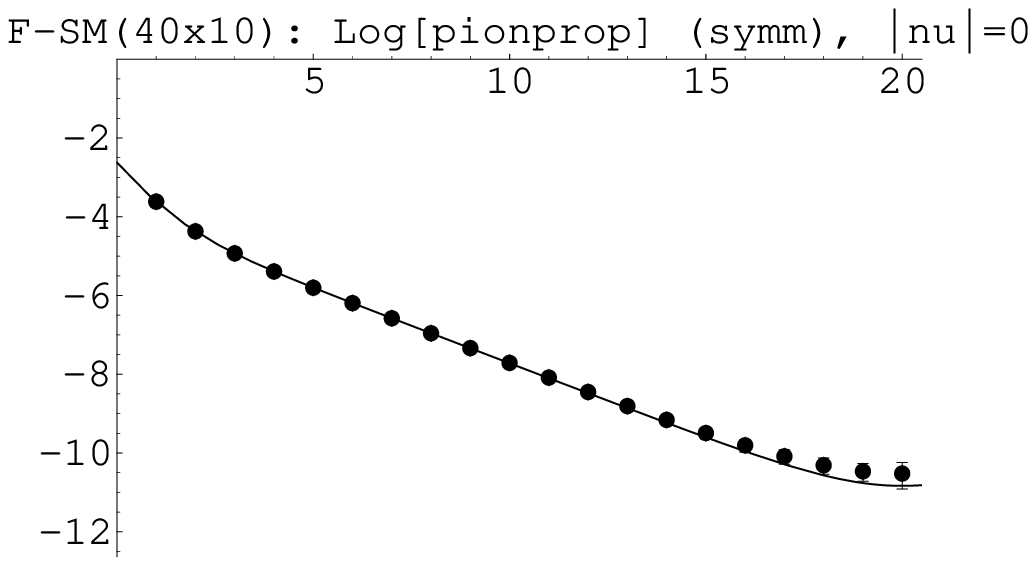,width=5.2cm,height=3.0cm}
\\
\epsfig{file=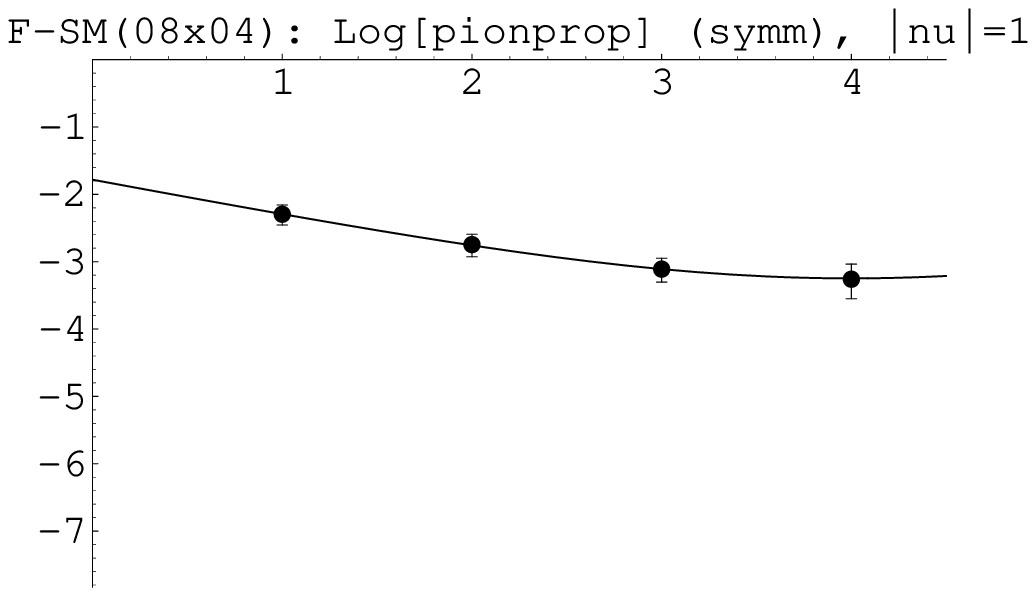,width=5.2cm,height=3.0cm}
\hspace*{0.1mm}
\epsfig{file=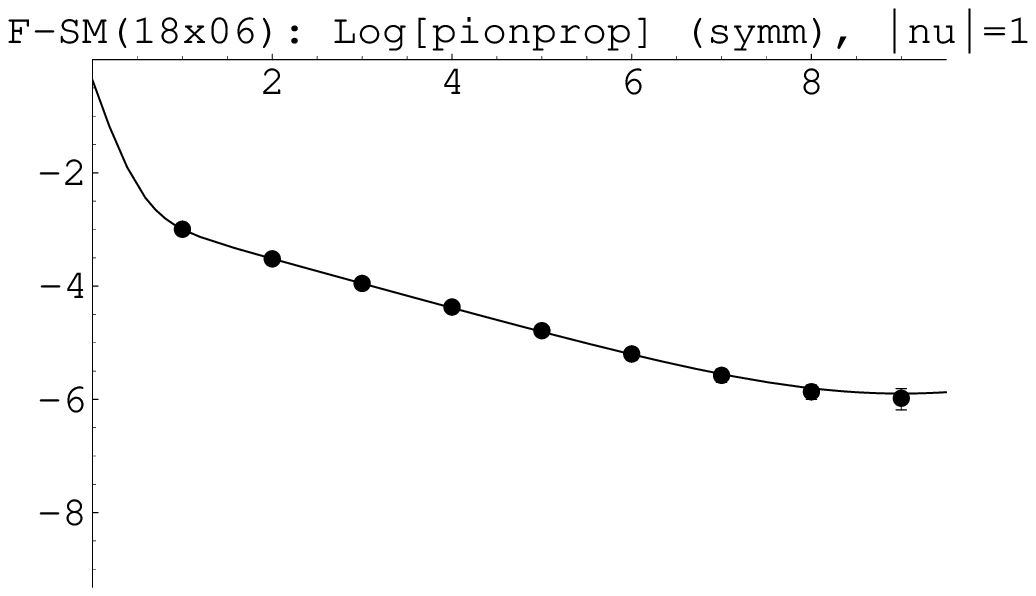,width=5.2cm,height=3.0cm}
\hspace*{0.1mm}
\epsfig{file=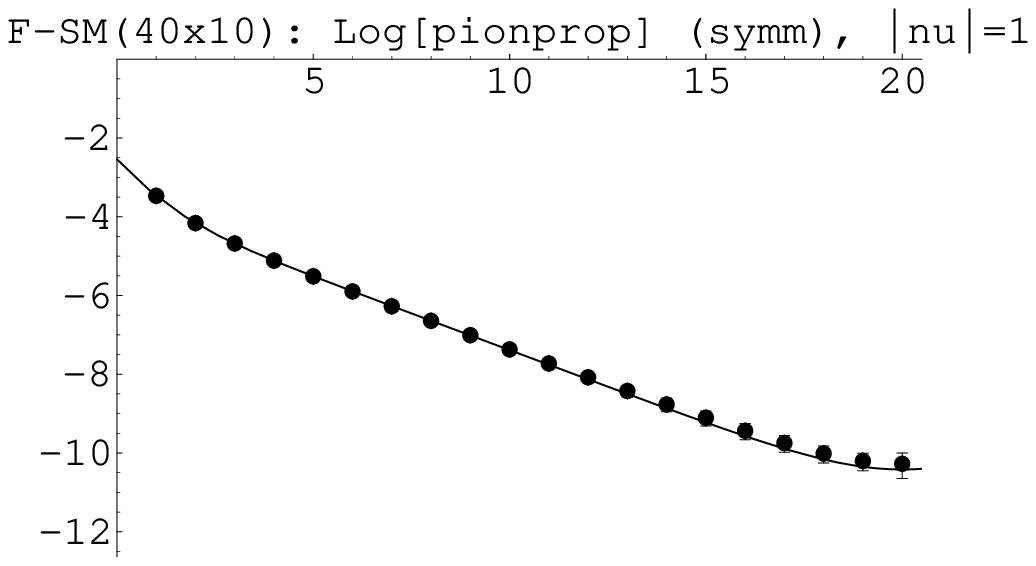,width=5.2cm,height=3.0cm}
\\
\hspace*{5.2cm}
\hspace*{0.1mm}
\epsfig{file=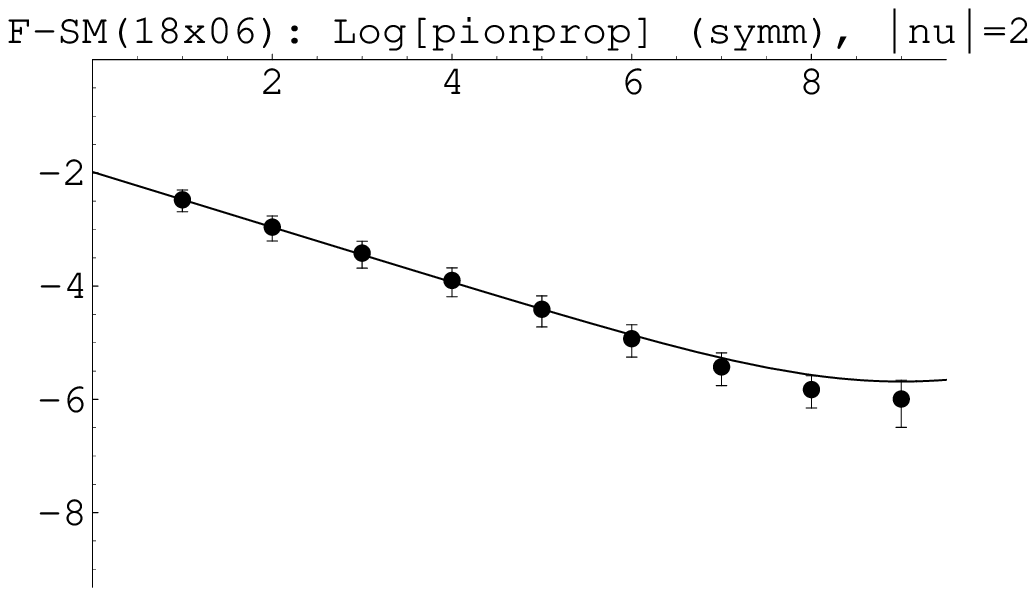,width=5.2cm,height=3.0cm}
\hspace*{0.1mm}
\epsfig{file=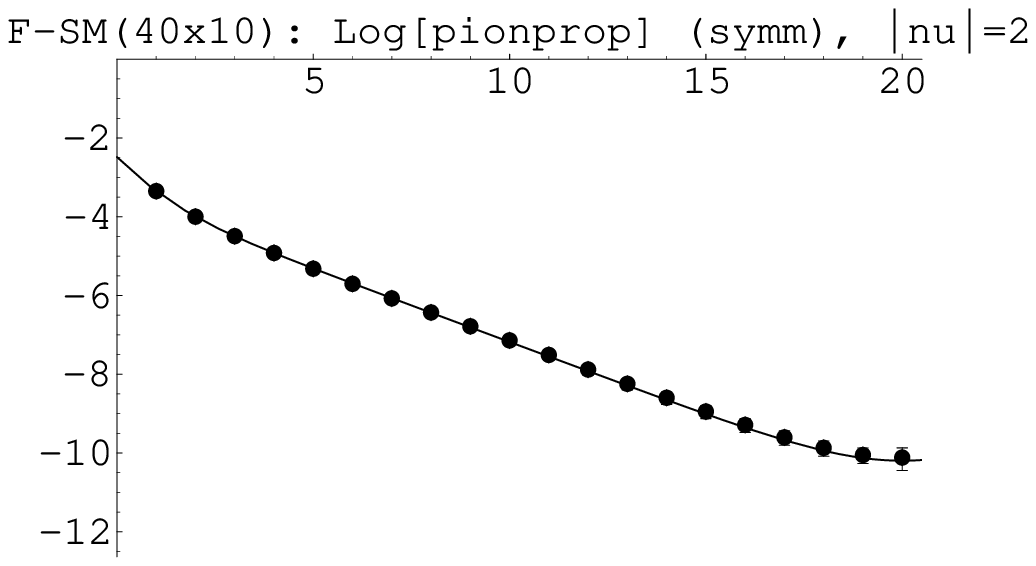,width=5.2cm,height=3.0cm}
\\
\hspace*{5.2cm}
\hspace*{0.1mm}
\hspace*{5.2cm}
\hspace*{0.1mm}
\epsfig{file=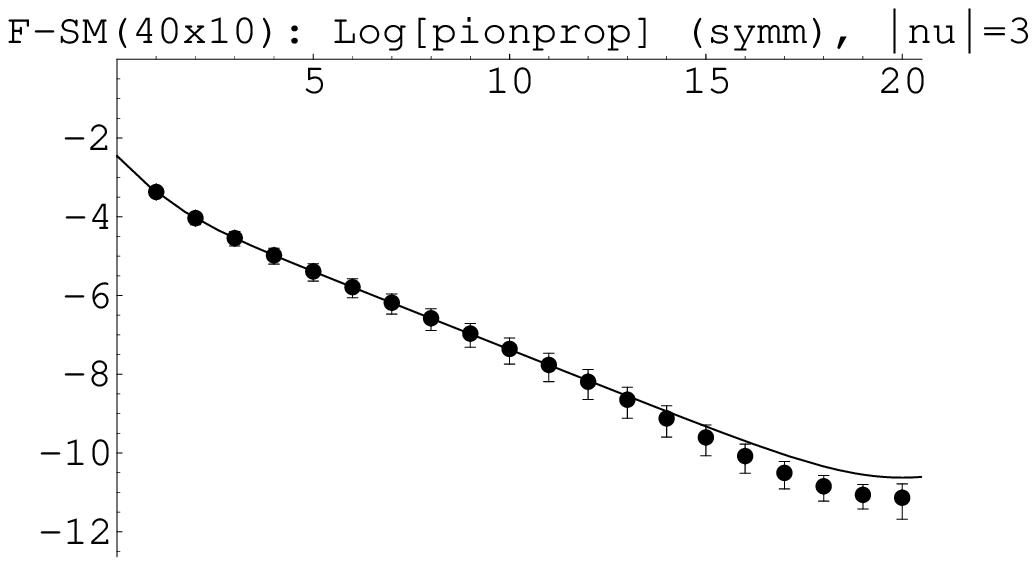,width=5.2cm,height=3.0cm}
\vspace*{-13mm}
\caption{Logs of the sectoral pseudoscalar two-point correlators in the small
(left), intermediate (middle) and large (right) Leutwyler-Smilga regime,
together with fits to the sum of two cosh-functions.
The secto- ral dependence of the effective pion mass disappears as one goes
from the small-$x$ to the large-$x$ regime.
\vspace*{-2mm}}
\end{figure*}

The final step is to evaluate the pseudoscalar two-point correlator
\beq
\<(\bar u\gamma_5 d)(0) (\bar d\gamma_5 u)(x)\>
\label{corr}
\eeq
at spacelike separations (note that in the chiral limit the multiflavour
Schwinger model shows a second order phase-transition with $T_{\!c}\!=\!0$
\cite{SMphase}, hence the rectangular shape of the manifolds).
For practical reasons, I have built the propagator from Wilson fermions, even
though there is an imminent risk that this `hybrid' formulation (sea-quarks
staggered, current-quarks Wilson) suffers from serious field-theoretic problems
(e.g.\ it's not clear whether there exists a bounded, symmetric, positive
transfer matrix \cite{Sommer}).
To avoid a `partially quenched' situation, $\ka$ is chosen such that
${1\ovr2}(\kappa^{-1}\!-\!\kappa_\mathrm{crit}^{-1})\!\simeq\!m$.
The result for the correlators is shown in Fig.\ 2, together with fits
to the sum of two cosh-functions.

As expected, a {\em pronounced sectoral dependence\/} of the pseudoscalar
propagator shows up in the {\em small\/}%
\footnote{Assuming universality in $x$, it seems that for $m\!\to\!0$ at fixed
$V$ the dominant long-range contribution to (\ref{corr}) stems from
$\nu\!=\!\pm1$, and this is at variance with what happens if one works directly
at $m\!=\!0$ in a finite volume, since there only $\nu\!=\!0,\pm2$ contribute
\cite{Azakov:2001pz}, i.e.\ our findings support previous skepticism about the
mass perturbation approach in the Schwinger model with $\Nf\!\geq\!2$
\cite{AntiMassPert}.}
LS regime.
For $x\!\simeq\!1$ there is a remnant sensitivity; the sectoral pion masses as
extracted from the fits still have a tendency to {\em decrease with $|\nu|$},
but none of them is far from the physical value.
In the {\em large\/} LS regime ($x\!\gg\!1$) the sectoral correlators (and
hence the sectoral pion masses) {\em agree surprisingly well\/} with each
other.
In summary, our findings {\em confirm\/} the LS prediction
\cite{Leutwyler:1992yt}, even though their `auxiliary' condition at large $x$
has been {\em reversed\/} as to guarantee that the pion would fit into the box
(what is the usual situation on the lattice).
It would be interesting to see this type of investigation repeated in full QCD.


\end{document}